\documentclass[aps, prl,twocolumn,groupedaddress]{revtex4-2}
\usepackage{CJK}
\usepackage{color}
\usepackage{graphicx}
\usepackage{dcolumn}
\usepackage{subfigure}
\usepackage{graphics}
\usepackage{amsmath}
\usepackage{amsfonts}
\usepackage{amssymb}
\usepackage{epsf}
\usepackage{bm}
\usepackage{ulem}
\usepackage{makecell} 
\usepackage{boldline} 

\usepackage[labelsep=period]{caption}
\begin{document}
\begin{CJK*}{GB}{ } 

\title{Assessing the Impact of Metacognitive Post-Reflection Exercises on Problem-Solving Skillfulness}
\author{Aaron Reinhard, Alex Felleson, Paula C. Turner, and Maxwell Green}

\address{
Department of Physics, Kenyon College, 201 North College Rd., Gambier, Ohio 43022, USA
}

\date{\today}
\begin{abstract}
We studied the impact of metacognitive reflections on recently completed work as a way to improve the retention of newly-learned problem-solving techniques.  Students video-recorded themselves talking through problems immediately after finishing them, completed ongoing problem-solving strategy maps or problem-sorting exercises, and filled out detailed exam wrappers.  We assessed students' problem-solving skillfulness using a combination of validated instruments and final exam questions scored using a rubric that targets problem-solving best practices. We found a small but significant difference between the rubric score distributions for the control and treatment groups. However, a multiple ordered logistic regression using treatment and Force Concept Inventory (FCI) pre-test score as predictors showed that this difference is better explained by the latter. The surprising impact of conceptual preparation on problem-solving skill suggests two things: the importance of remediation for students with insufficient conceptual understanding and the need to consider problem-solving interventions in the context of students' conceptual knowledge base.
\end{abstract}

\maketitle		
\end{CJK*}

\section{Introduction}

Two learning goals are ubiquitous in an introductory physics course: the ability to reason qualitatively using physics and the development of quantitative and problem-solving skills. The most widely-used strategies to address the former goal have been active learning techniques~\cite{Freeman,Mazur,Garvin,SokoloffandThornton,Shaffer,Akcayir,Miller}.  These strategies are well-studied~\cite{Freeman}; produce dramatically improved attitudes, course performance, and normalized gains on concept inventories~\cite{Hake,Crouch,Prince,PrinceandFelder,Freeman}; and disproportionately benefit students from underrepresented groups~\cite{Haak,Theobald}.  The development of quantitative and problem-solving skills, on the other hand, has received less attention in the physics education research literature, although that has begun to change in recent years.

Historically, research  on problem solving in physics has focused on the approaches of novice vs. expert problem solvers.  For example, novices tend towards haphazard formula-seeking, pattern matching, and premature jumps to quantitative expressions~\cite{Reif,Larkin,VanHuelven,Mazur}.  Experts, on the other hand, organize information after reading a problem~\cite{ReifandHeller}, plan their solutions before beginning~\cite{Larkin}, classify problems according to major principles~\cite{Mestre}, and move between multiple representations~\cite{Kohl}.  In short, they access higher-order organizational structures to reduce the cognitive load on their short term memory~\cite{Chi,Sweller,Brown}.  However, the body of literature on \textit{pedagogical strategies} aimed at moving student behaviors from novice to expert is brief but growing.

Metacognition has emerged as a fruitful approach to teaching problem-solving skills.  Metacognition, or thinking about one's own thinking~\cite{Flavell}, is considered to be an essential element of the learning process~\cite{Pintrich,Georghiades,Rickey}.  Explicit focus on metacognition has been shown to improve student motivation and self-regulation~\cite{Zepeda}.  Research has shown that students who demonstrate a variety of metacognitive skills perform better on general problem solving tasks~\cite{Rozencwajg,Yuruk,Zohar}.

Several strategies that use metacognition to improve problem-solving skills have been described in the literature.  Lewis showed that teaching students to graphically represent arithmetic word problems produced significant gains in problem-solving performance~\cite{Lewis}.  Multiple studies have focused on teaching explicit strategies for problem solving and reinforcing these skills through collaborative group work~\cite{Heller,Heller2,SandiUrena,Mota}. Other work has demonstrated the benefit of verbalizing one's thought process while engaged in the act of solving a problem~\cite{Mayer,Pressley,Steif}.  In their meta-analysis of 178 peer-reviewed studies on metacognition in science education, Zohar and Barzilai note that ``the most prominent practice is the use of metacognitive cues and prompts in the course of instruction''~\cite{Zohar}.

In this work, we present a systematic program of metacognitive activities designed to improve problem-solving skillfulness in an introductory mechanics course.  However, unlike most recent work, we focus not on the acquisition of new skills, but on the retention of recently acquired skills.  We have noticed that students can sometimes perform a difficult task in class or on homework with minor assistance, but fail to reproduce this work on an exam.  It is as if the learning is not fully assimilated.  By taking regular opportunities to reflect on new learning just after it has occurred, we hope to solidify dimly-grasped concepts and make gains more permanent.  

\section{Motivation}

One of us, AR, taught a two-semester, algebra-based introductory physics sequence for students in the life sciences over four consecutive years at Otterbein University.  As suggested above, one goal was to build strong conceptual understanding of the principles of physics.  Thus, the classes featured active learning strategies such as Peer Instruction~\cite{Mazur}, interactive lecture demonstrations~\cite{SokoloffandThornton}, and guided-inquiry exercises~\cite{Shaffer,Knight}.  In the mechanics course, we administered the Force Concept Inventory (FCI) as a pre- and post-test and measured the class-averaged normalized gain.  The gains were consistent with published norms for active-learning classrooms~\cite{Mazur,Hake}.

A complementary goal was the development of strong problem solving and quantitative reasoning skills. As such, the instructor developed a rigorous and structured approach to teaching problem-solving best practices.  The pedagogical sequence went as follows. First, the instructor introduced the strategy for a new type of problem, like Newton's second law or conservation of energy, by doing sample problems at the board.  He modeled good problem-solving technique and talked through his reasoning in real time.  During this task, the instructor would sometimes pause so students could discuss important steps or strategies with their peers.  Immediately following this exercise, students worked in teams on similar problems using a structured worksheet.  They were required to focus on important steps like extracting the given parameters and the needed variable from text, drawing a diagram with initial and final states, constructing a useful free body diagram, labeling the position of zero potential energy, or setting up the appropriate equations without numbers.  Finally, the instructor would assign similar, but more challenging, homework problems to help students master these new skills.  Student solutions were submitted on paper and hand-graded, with points explicitly assigned to the important solution elements required for each problem type.

The quality of the students' written work was very high.  We attribute this, in part, to the rigorous and scaffolded approach to teaching problem solving.  However, despite the students' good performance, we noticed a problem.  They would sometimes complete a challenging homework problem with minor assistance, but were unable to duplicate the task on a test or future assignment.  It was especially jarring when a student would choose an inappropriate strategy for a specific type of problem---for example, using a kinematic equation on a problem involving the collision between two objects.  It seemed as if the problem-specific learning gains students achieved on the homework were transient and quickly lost.  Students seemed to be acquiring the requisite skills to solve challenging problems, but these skills were not permanent. It was apparent that a new approach to reinforcing problem-solving skills was needed.

\section{Experimental Design}

We attempted to address this lack of skill retention by focusing on metacognition.  Specifically, we considered the impact of metacognitive reflection on work \textit{already done} by students or on material \textit{just learned}. We hoped to teach our students to use post-learning reflection to make gains permanent and to fit their newly acquired knowledge into the coherent, organized structures characteristic of expert learners~\cite{Chi,Sweller,Brown}.  Our idea was to promote metacognitive post-reflection on problem solving at three ``magnitude scales'' relevant to a semester-long introductory physics course: the single-problem level, the single-unit level, and the midterm-exam level.

\begin{center}
\begin{figure}[htp]
\centerline{ \scalebox{0.9} {\includegraphics{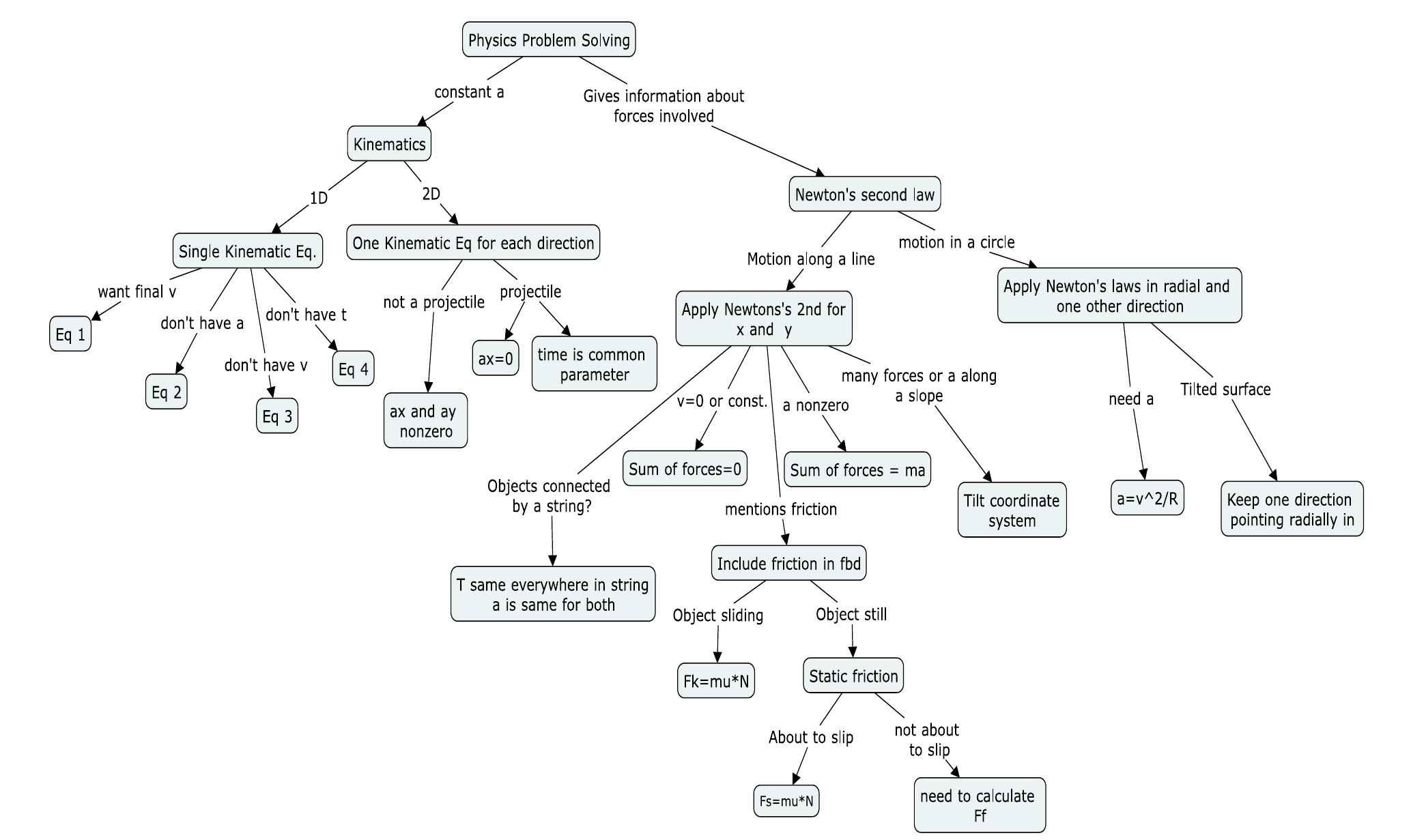}}} \caption{ \label{Fig0} Example problem-solving strategy map generated at the end of the units on 1D kinematics, 2D kinematics, and Newton's Laws.  The instructor created this map and shared it with the students to demonstrate the intended format.}
\end{figure}
\end{center}

The three items of our plan are summarized as follows:

\begin{enumerate}
  \item \textbf{Video recorded ``think-alouds''}: Students would use Flipgrid~\cite{Flipgrid} to video-record themselves talking through one finished homework problem each week, immediately after completing that problem.  Students were told to state their thinking at every step of the problem, in the same way as the instructor when presenting a sample problem in class. This differs from other studies, where students verbalized their thought process while solving the problem for the first time~\cite{Mayer,Pressley,Steif}.  The idea was to encourage students to reflect on how they completed a problem and to make their thought processes explicit by stating them aloud. We wanted students to consider the structures and strategies that were effective in solving a problem, \textit{just} when those structures and strategies were hazy and partially grasped. The videos were graded for completion; however, the students also received qualitative feedback from the instructor or a TA.  After the instructor approved the videos, every student in the class had access to all videos of their peers to use as a study aid.

  \item \textbf{Problem-solving strategy maps}: At the end of each unit,  students would update an ongoing problem-solving strategy map~\cite{note}.  We intended students to use the map to fit the new topic into the framework of material they had already learned that term.  The strategy maps were intended to follow best practices for concept maps: a top-down, hierarchical organization of ideas needed to solve physics problems~\cite{Novak}. We hoped that the maps would consist of things which, although obvious to experts, are not typically obvious to beginners: conservation of mechanical energy is best used when objects change height, conservation of linear momentum is best used when things collide, etc. A student could use this map as a ``decision-making tree'' when encountering a new problem of unknown type.  Research has shown that expert learners employ sophisticated mental structures to fit new information into the context of what they already know~\cite{Chi,Sweller,Brown}. We hoped that this map would be a physical manifestation of the ongoing mental framework students construct as they learn. The students digitally managed their maps~\cite{cmap} to streamline the process of updating, improving, and modifying them, as well as collecting instructor feedback.  An example of the intended format is shown in Fig.~\ref{Fig0}.
  
  \item \textbf{Exam wrappers}: We required students to complete detailed exam wrappers after taking each midterm exam.   Exam wrappers are short surveys where students reflect on their study strategies and on their exam performance, with an eye towards targeted improvements~\cite{Kaplan}. We wanted students to critically examine the places they succeeded and the places they struggled, and to correlate those outcomes with problem-solving strategies, study habits, and general behavior in the course.  We expected each student to gain insight into the unique ways in which they learn. By completing this exercise, we hoped to demonstrate to students the correlation between their use of best practices and success on exams.
\end{enumerate}

\onecolumngrid
\twocolumngrid

We administered three assessment instruments. The first was the Metacognitive Activities Inventory (MCAI), ``an instrument designed to specifically assess students' metacognitive skillfulness during problem solving''~\cite{Cooper}.  Students completed this inventory online, outside of class, both at the beginning and the end of the intervention semesters. The other two instruments were the Force Concept Inventory (FCI) and the Mechanics Baseline Test (MBT)~\cite{Wells,Hake,Hestenes}.  We administered the FCI as a pre- and post-test and the MBT as a post-test only, all  during class time. A nominal amount of extra credit was given for the post-tests.   Both instruments test conceptual understanding of physics, but the MBT requires moderate amounts of computation and quantitative reasoning~\cite{Hestenes}. We used the correlation between FCI and MBT scores as a metric for how well our students used quantitative reasoning to understand ideas in physics. There is published data on correlations between FCI scores and MBT scores~\cite{Hestenes}. A student scoring higher on the MBT relative to their FCI post-score would indicate strong quantitative and problem-solving skills, given their level of conceptual understanding as measured by the FCI.

\begin{center}
\begin{figure}[htp]
\centerline{ \scalebox{1.1} {\includegraphics{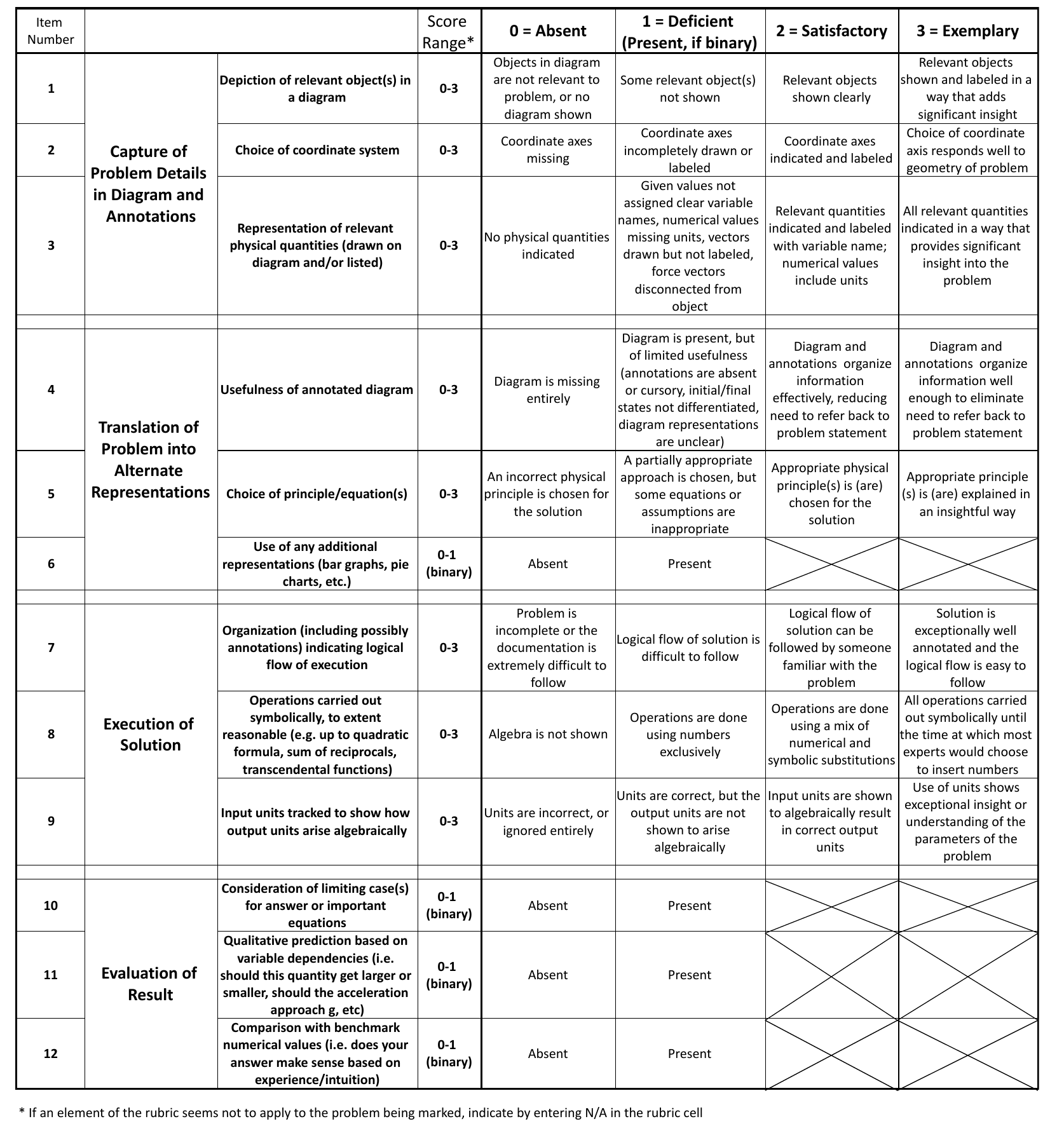}}} \caption{ \label{Fig1} Rubric used to assess student use of problem-solving best practices.}
\end{figure}
\end{center}

We also performed a controlled study of students' use of expert problem solving strategies on final exam problems. We collected one year of final exam solutions from students who did not receive the intervention.  We then assigned the same problems on future exams, in courses where the intervention was employed.  External evaluators scored these problems, using a rubric we developed, to assesses student use of problem-solving best practices.  While the rubric is new, it was created to mirror the aspects of expert-like problem solutions enumerated in the literature, including planning a solution, use of alternate representations, skilful execution, and evaluation of the result~\cite{Larkin,Reif,Mestre,Kohl}. The rubric is shown in Fig.~\ref{Fig1}. 

\onecolumngrid
\twocolumngrid

\section{Implementation and Qualitative Results}

We implemented our study in an introductory calculus-based mechanics course taught by AR at Kenyon College during the fall semesters 2017, 2018, and 2019. The fall 2017 class did not receive an intervention, but we collected scores on the FCI and MBT, along with responses to the final exam questions.  The fall 2018 class received the first year of the intervention.  Starting with the second homework assignment, students video-recorded themselves talking through one problem each week.  This problem was chosen to be representative of the ``type'' for that homework and was of medium difficulty.  Starting with the third assignment, students turned in a short list of elements that each problem on that homework had in common.  This list might include common solution strategies, conditions that must be true to use those strategies, or common problem features.  Then, each Friday, the students would use this list to update their problem-solving strategy map with features (bubbles) and connections (arrows) helping them navigate the decisions necessary to solve this new type of problem (see Fig.~\ref{Fig0}).  Finally, after each of the two midterm exams, students filled out a nine-question exam wrapper.  The instructor reviewed these responses and returned them to the students in the week before the next exam, to remind students of their impressions regarding their work on the previous exam.

We collected anecdotal evidence about the success of each of these interventions during the semester. We define success as high levels of student enthusiasm for and adequate performance on each intervention.   Many of our impressions were supported by student responses to course evaluation questions.  The think-aloud videos were very successful.  After a few weeks, the quality of the videos was uniformly high.  This was the most popular intervention among students, and they appreciated that it forced them to verbalize their thinking.  We also had the impression that the exam wrappers were a useful exercise.  The student responses were thoughtful, and a number of students told us that they changed their habits as a result of considering correlations between their behavior and their exam performance.

In contrast, the problem-solving strategy maps were not successful.  Students were given 15-20 minutes of class time each Friday to discuss in pairs while working on individual maps.  Before the first Friday session, students watched a video on concept maps and answered several questions about their purpose and execution.  The instructor used class time several times throughout the semester to discuss elements of a good problem-solving map.  He showed examples of strong and weak maps, and students worked in groups to use each sample map to make decisions about the strategy for new problems of unknown type.

The interactions among students were relatively sparse during in-class map creation time, and the quality of many maps never rose to a useful level.  Maps were often disjointed, lacking the salient characteristics of each problem type, and contained weak, confusing, or circular links between characteristics. Despite repeated instruction on the purpose of the maps and examples of good and bad maps, the students were unable to make broad connections while updating the maps in real time (i.e. \textit{just} after learning a concept).  On the course evaluation, students admitted to rarely using their maps when not required to.

We determined that this exercise was not developmentally appropriate for students in an introductory physics course.  It was too much to expect students to be able to articulate the connections between strategies as they learned them each week.  Constructing the maps required students to reason deductively, from the abstract to the concrete, a characteristic of experts~\cite{Brown}.  We therefore hypothesized that a task which required inductive reasoning, or working from the concrete to the abstract, would fall more within our students' zone of proximal development.

In fall 2019, or the second year of our intervention, we changed the exercise that targeted the single-unit level.  We again administered the think-aloud videos and the exam wrappers; however, we replaced the problem-solving strategy maps with a problem-sorting exercise.  Three times throughout the semester students were put into groups and presented with about 10 new problems of various types, in random order.  Groups were asked to identify the type of problem each question represented, list the features of the problem which led to this identification, use these lists to identify common features and solution strategies for each category of problem, and take turns setting up each problem on a whiteboard to the point where the solution would involve only math to complete.  Once they could say, for example, ``plug equation 2 into equation 1 and solve for $a$,'' they were instructed to stop and set up another problem.

We observed much more enthusiastic conversations during this exercise, as compared with the discussions during the composition of problem-solving strategy maps.  These interactions often featured rich debate and occurred at higher levels on Bloom's taxonomy~\cite{Bloom}.  On the post-semester course evaluation, students chose this exercise as frequently as the think-aloud videos as their favorite of the metacognitive exercises.  In future years, we plan to implement this exercise more than three times per semester.

\section{Results}

\vspace{-0.6cm} 

\begin{table}[htp]
\caption{Class-average percentage scores (reported with margins of error at 95\% confidence) on standard PER instruments. Pairwise deletion was used so that pre- and post-test averages for the same assay represent the same subset of students.}  
\label{tab:PERassays}
\def\arraystretch{1.3} 
\resizebox{\columnwidth}{!}{ 
\begin{tabular}{|c|c|c|c|c|c|c|} 
\hline
cohort & enrollment & \makecell{FCI pre\\(\%)} & \makecell{FCI post\\(\%)} & \makecell{MBT post\\(\%)} & MCAI pre (\%) & MCAI post (\%) \\ \Xhline{2\arrayrulewidth} 
1 & 24 & $59 \pm 12$  & $78 \pm 7$ & $63 \pm 8$ & not measured  & not measured \\ \hline
2 & 19 & $62 \pm 15$  & $85 \pm 7$ & $64 \pm 7$ & $82 \pm 4$    & $83 \pm 4$ \\ \hline
3 & 24 & $70 \pm 10$ & $84 \pm 7$ & $65 \pm 7$ & $79 \pm 5$ & $82 \pm 4$ \\ \hline
\end{tabular}
} 
\end{table}

Table~\ref{tab:PERassays} shows class-average percentage scores on three assays used to characterize conceptual understanding, problem-solving ability, and metacognitive attitudes of students (the FCI, MBT, and MCAI, respectively). Cohorts 1, 2, and 3 refer to students from Fall 2017, 2018, and 2019, respectively. Thus cohort 1 experienced no metacognitive intervention, while cohorts 2 and 3 participated in the interventions as described above.

When converted to normalized gains (using gain = ($\langle$post$\rangle$ - $\langle$pre$\rangle$)/(1- $\langle$pre$\rangle$)), all three cohorts show high FCI gains (46\%, 61\%, and 47\%, respectively), indicative of an effective active-engagement course~\cite{Mazur,Hake}. For each cohort, the MBT averages trail the FCI post-instruction averages by 15, 21, and 19 percentage points, respectively, which is in the range of typical performance when these metrics are compared~\cite{Hestenes}. The variation between cohorts in these differences is almost entirely due to differences in the FCI post-instruction scores, since the MBT percentage scores are essentially the same across cohorts. Attitudes toward metacognitive activities hold steady through the semester for cohorts 2 and 3, which is typical in MCAI percentages reported by~\cite{Cooper} for pre- and post-course assessments using the MCAI. 

The whole-course averages on the assays presented in Table~\ref{tab:PERassays} do not indicate clear differences between the treatment groups (cohorts 2 and 3) and the control group (cohort 1). We therefore considered assessments of student problem-solving techniques made by three independent raters using the rubric presented in Fig.~\ref{Fig1}. One characteristic (item 6, use of any additional representations) was so infrequently scored as present by any rater that it was excluded from subsequent analysis. An intraclass correlation (ICC) test of inter-rater reliability for aggregated rubric data showed good agreement between raters ($\kappa$ = 0.84, p-value = 0). Similar ICC tests on rubric data disaggregated by rubric characteristic showed that the non-binary rubric items (1-5, 7-9) each had moderate to good agreement ($\kappa$ = 0.42 to 0.87), as shown in Table~\ref{tab:IRR}. 


\begin{table}[htp]
\caption{Inter-rater reliability for rubric scores}
\label{tab:IRR}
\resizebox{\columnwidth}{!}{ 
\begin{tabular}{|l l V{2} c|} 
\hline
\multicolumn{2}{|c V{2}}{Rubric Item}                 & $\kappa$\\ 
\Xhline{2\arrayrulewidth} 
1   & Depiction of relevant objects             & 0.71\\ \hline
2   & Choice of coordinate system               & 0.87\\ \hline
3   & List of quantities                        & 0.42\\ \hline
4   & Usefulness of diagram                     & 0.57\\ \hline
5   & Choice of principle                       & 0.67\\ \hline
6   & Additional representations                & \makecell{excluded from analysis} \\ \hline
7   & Organization and flow of execution        & 0.54\\ \hline
8   & Operations performed symbolically         & 0.60\\ \hline
9   & Input units tracked algebraically         & 0.65\\ \hline
10  & Limiting cases analyzed (binary)          & 0.07\\ \hline
11  & Qualitative predictions made (binary)     & 0.07\\ \hline
12  & Comparison with benchmark values (binary) & 0.24\\ \hline
\multicolumn{2}{|l V{2}}{All rubric items aggregated} & 0.84\\ \hline
\end{tabular}
}
\end{table}

\begin{figure}[htp]
\centerline{ \scalebox{0.4} {\includegraphics{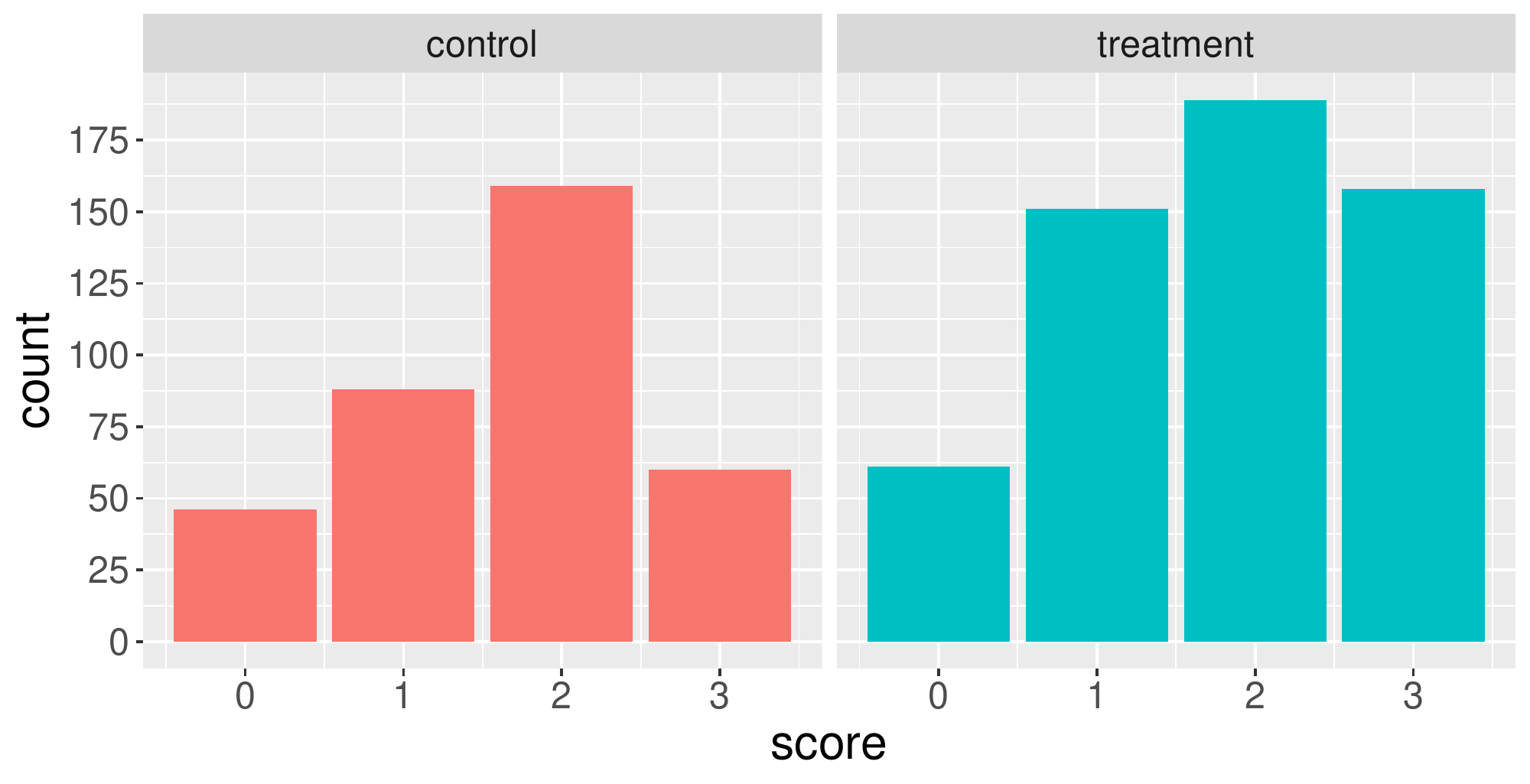}}} \caption{ \label{rubric score distribns} Distribution of rubric scores for control and treatment groups.}
\end{figure}

We first examined the distributions of rubric scores for the treatment and control groups, shown in Fig.~\ref{rubric score distribns}.  We focused on rubric items with good inter-rater reliability, which required us to exclude the binary questions. A Mann-Whitney test, assuming the two distributions have the same general shape, showed a significant difference between their median ranks ($W$ = 90948, p-value = 0.037), although with a small effect size (r = 0.07, 95\% CI [0.01, 0.14]). This indicates a small but nonzero  difference (95\% CI [-4.63e-5, -1.67e-5]) in the use of best practices when solving problems between the two groups. When compared for a single rubric characteristic at a time, the distributions did not differ.

An ordered logistic regression run on all scores given for non-binary rubric items agreed with this result: being in the treatment group increased the ordered log odds of a student earning a higher score by 0.260 (p-value $= 0.037$, 95\% CI [0.02, 0.50]). This analysis was run on the same set of students as the model shown in Table~\ref{tab:OLOG2}---that is, students who took the FCI pre-test---to allow for reasonable comparison between the two. Parameter estimates from the model are given in Table~\ref{tab:OLOG1}.

\begin{table}[htp]
\caption{Results of ordered logistic regression model of rubric score versus presence in the treatment group.}
\label{tab:OLOG1}
\centering
\begin{tabular}{r|rrrr}
  \hline
 Variable \hspace{1.1cm} & \hspace{0.1cm} Coefficient & \hspace{0.1cm} Std. Error & \hspace{0.1cm} t value & \hspace{0.1cm} p value \\ 
  \hline
TreatmentGroup & 0.260 & 0.125 & 2.082 & 0.037 \\ 
Intercepts: \hspace{0.15cm} 0$|$1 & -1.848 & 0.127 & -14.558 & 0.000 \\ 
  1$|$2 & -0.312 & 0.101 & -3.093 & 0.002 \\ 
  2$|$3 & 1.322 & 0.111 & 11.954 & 0.000 \\ 
   \hline
\end{tabular}
\end{table}


The odds ratio for the treatment group effect is calculated as $\exp{(\mbox{coefficient})}$---here, $\exp{(0.260)}$---and comes out to be 1.30 (95\% CI [1.02, 1.66]). This means that students in the treatment group (cohorts 2 and 3) had 1.30 times higher odds of getting a higher rubric score (e.g., 3 or 2 rather than 1) compared to students in the control group when FCI pre-test scores are held constant.

Due to the quasi-experimental design of our study, we turned to multiple regression techniques with FCI pre-score as the first predictor variable to control for pre-existing differences in physics knowledge between the control and treatment groups before looking for a treatment effect. Results from an ordered logistic regression model on the effect of FCI pre-score \textit{and} membership in the treatment group (Table~\ref{tab:OLOG2}) show that FCI pre-score is a much more significant predictor of rubric score ($\beta_{FCI pre}=0.03$, p-value=0.00, 95\% CI [0.01,0.05]), and membership in the treatment group is no longer a significant predictor after FCI pre-score is taken into account ($\beta_{treatment}=0.16$, p-value=0.20, 95\% CI [-0.09,0.41]). This indicates that when we control for pre-existing differences in students' exposure to concepts in mechanics, we can no longer claim a significant effect for our intervention.

\begin{table}[ht]
\caption{Results of ordered logistic regression model of rubric score versus FCI pre-score and group}
\label{tab:OLOG2}
\centering
\begin{tabular}{l|rrrr}
  \hline
 Variable \hspace{1.1cm} & \hspace{0.1cm} Coefficient & \hspace{0.1cm} Std. Error & \hspace{0.1cm} t value & \hspace{0.1cm} p value \\ 
  \hline
FCI pre-score & 0.030 & 0.008 & 3.705 & 0.000 \\ 
  TreatmentGroup & 0.162 & 0.128 & 1.269 & 0.204 \\ 
  Intercepts: \hspace{0.15cm} 0$|$1 & -1.344 & 0.184 & -7.284 & 0.000 \\ 
  \hspace{1.83cm} 1$|$2 & 0.210 & 0.173 & 1.212 & 0.225 \\ 
  \hspace{1.83cm} 2$|$3 & 1.863 & 0.185 & 10.098 & 0.000 \\ 
   \hline
\end{tabular}
\end{table}

The odds ratio calculated from the treatment term's coefficient is 1.18 (95\% CI [0.92, 1.51]). That means that students who were in the treatment group had odds of getting a higher rubric score that were 1.18 times that of the students in the control group, when holding FCI pre-test scores constant. By comparison, log-odds ratio associated with the FCI pre-score term is 1.03 (95\% CI [1.01, 1.05]).

We were unable to make a model controlling for the effect of students' pre-existing tendencies toward employing metacognitive techniques as measured by the MCAI because we did not collect MCAI pre-scores for students in cohort 1, and we were unable to find another variable in our data set with a strong enough correlation to MCAI pre-score for cohorts 2 and 3 that we could use it as a proxy.

\section{Discussion}

Our metacongnitive intervention did not yield statistically significant differences on assays that measure problem-solving skill.  We can, however, draw several important conclusions.  First, minor changes in the execution of this study might yield significant results.  We employed a quasi-experimental design, in which students were not randomly sorted into intervention and control groups.  This was done because of the low $n$ typical of introductory physics courses at a liberal arts college (we had $n=25$, 19, and 25 in our three cohorts, respectively).  Low $n$ is likely the reason we don't see statistically significant results.  The intervention vs. no-intervention histograms in Fig.~\ref{rubric score distribns} are, in fact, statistically different, and the ordered logistic regression run on rubric score with treatment as the only explanatory variable concurs. But the ordered logistic regression model does not show a significant difference when FCI pre-test score is included as a predictor, due to the large standard error on the regression coefficient estimate for the treatment variable, as shown by its wide 95\% confidence interval.  It is possible that a data set with a larger $n$ might show the treatment having a significant effect along with FCI pre-score. Were that the case, we would probably also see similar odds ratios to the ones calculated from Table~\ref{tab:OLOG2}. Since the odds ratio attributed to a variable is the best representation of its relative effect size, such a result would mean that while both variables are significant predictors of rubric score, presence in the treatment group has an obviously larger effect on score, which is an encouraging possibility. In other words, comparing the two predictors' effect sizes (via odds ratio) shows that FCI pre-score overshadows treatment in significance only, not magnitude of effect, and lack of significance may be due to our small sample size rather than a true lack of effectiveness of our treatment. 

To yield the clearest results, our intervention should be implemented in a large introductory classroom in which students are randomly selected to participate in the intervention activities.  Barring this, one could implement the intervention in some, but not all, sections of a multi-section, single-year course.

The two cohorts that received the intervention showed little change in score on the MCAI over the course of the semester~\cite{Cooper}.  However, it would be interesting to administer the MCAI to both the control and treatment groups as a pre- and post-test.   This would allow us to determine what effect our intervention has on individual students' MCAI score while controlling for other factors. For example, it could be that the intervention is correlated with higher MCAI gains for students who come to the course with a low FCI pre-test score, even though the class average MCAI score does not change.

The most significant result is that the FCI pre-test score is the dominant predictor of use of metacognitive strategies (rubric score) on final exam responses.  The FCI is intended to assess qualitative understanding of the principles of Newtonian mechanics~\cite{Hestenes}, a topic which makes up only a fraction of the content of a first-semester course.  At first glance, is surprising that the FCI \textit{pre}-test would be a significant predictor of students' problem solving skillfulness \textit{after} a semester-long metacognitive intervention.

In their paper on how to interpret the FCI, Hestenes and Halloun state~\cite{Hestenes2}:

\begin{quotation}
  On the basis of our published evidence, we are confident in asserting that the FCI, coupled with a simple math test, is probably the most accurate predictor of performance in introductory physics in either high school or college.  Nevertheless, we advise against using it as a placement exam for such courses, because its high predictive power is indicative of inadequacies in the instruction rather than in the students.
\end{quotation}

Coletta \textit{et al.} show that the SAT is a good measure for students' general cognitive skills and even predicts a student's FCI gain~\cite{Mazur,Coletta}.  Note that cognitive skill is not correlated with the FCI pre-test score, but with the semester-long \textit{gain}.  This supports the idea that the FCI pre-test measures student preparedness.  Other studies show that students' grades in both math and physics courses are predicted by the FCI pre-test score~\cite{Henderson,Brum}.  Stoen \textit{et al.} generalize by stating ``FCI performance may reflect a number of student attributes including relational knowledge structures of physics concepts, expertlike attitudes, and problem-solving skills''~\cite{Stoen}.  The literature suggests that the FCI pre-test can be a very general predictor of success on quantitative tasks; however, it most likely indicates a student's level of preparation rather than their cognitive abilities.

This insight suggests several courses of action.  First, while Hestenes and Halloun are clear that the FCI should not be used as a placement test~\cite{Hestenes2}, we argue that it could be used as an assessment to guide remediation or targeted interventions. Examples might include additional tutoring support, arranging for students with different levels of preparedness to work together on group problem-solving tasks, or targeted feedback on homework assignments.  PT has recently experimented with ``first-step sheets.'' Students work collaboratively to compare, correct, and improve their individual attempts at setting up each problem on a homework assignment two days before it is due. The instructor circulates among groups, answering questions and clarifying issues.  One could use this class time to provide strategic support, or one could collect the sheets and give individualized feedback before the problems are due.  Such interventions might have an impact on both qualitative assessments as well as the development of problem-solving skills.  Second, because we find that student gains in problem-solving skill are so closely linked with a pre-test of qualitative knowledge, one cannot simply consider problem-solving interventions alone, in the absence of students' broader conceptual knowledge base.  Finally, our results might suggest a need to more broadly adopt evidence-based, active-learning techniques in K-12 science curricula, since insufficient conceptual preparation is correlated with a broad range of outcomes.

\section{Conclusion}

We have presented a study of the impact of metacognitive post-reflection exercises on problem solving skillfulness in an introductory physics classroom.  We assessed problem-solving skillfulness using a rubric designed to identify problem-solving best practices in students' written work on five final exam problems.  While the treatment and control groups do show statistically different rubric score distributions, this difference is no longer significant after controlling for FCI pre-test score. This surprising finding indicates that a lack of conceptual preparation has far-reaching effects and suggests that the FCI could be used as a tool to inform targeted interventions.  It also demonstrates the need to consider problem-solving interventions in the context of students' qualitative understanding. Given the encouraging difference in effect size between treatment and FCI pre-test score, it will be important to repeat this study with a larger sample size. This could either confirm our intervention's apparently large effect size or suggest unforeseen subtleties in the potential effectiveness of metacognitive interventions.

\textit{Acknowledgements.}  The authors acknowledge valuable input from Paul Wendel, both in the conception of the study and a careful review of the manuscript, and from Brad Hartlaub, for assistance with statistics.  This work was supported by NSF Grant PHY-1553179.


\begin{thebibliography}{99}
\bibitem{note} We define a ``unit'' as a period focused on a new problem-solving technique like Newton's laws or a conservation law.
\end{thebibliography}
\end{document}